\documentstyle[preprint,aps,epsf,floats]{revtex}

\begin{document}
\setlength\baselineskip{20pt}
\def\vslash{v\!\!\!\slash}
\newcommand{\nn}{\nonumber}

\preprint{\tighten \vbox{\hbox{CALT-68-2160} 
  \hbox{hep-ph/9803227} \hbox{} }}

\title{Extraction of the $D^*D\pi$ coupling from $D^*$ decays}
 
\author{Iain W.\ Stewart}

\address{California Institute of Technology, Pasadena, CA 91125  }

\maketitle

{\tighten 
\begin{abstract}

The decays $D^* \to D\pi$ and $D^* \to D\gamma$ are well described by heavy
meson chiral perturbation theory.  With the recent measurement of ${\cal
B}(D^{*+} \to D^+ \gamma$), the $D^{*0}$, $D^{*+}$, and $D_s^*$ branching
fractions can be used to extract the $D^*D\pi$ and $D^*D\gamma$ couplings $g$
and $\beta$.  The $D^* \to D\gamma$ decays receive important corrections at
order $\sqrt{m_q}$ and, from the heavy quark magnetic moment, at order $1/m_c$.
Here all the decay rates are computed to one-loop, to first order in $m_q$ and
$1/m_c$, including the effect of heavy meson mass splittings, and the
counterterms at order $m_q$.  A fit to the experimental data gives two possible
solutions, $g=0.27\,^{+.04}_{-.02} \,^{+.05}_{-.02}$, $\beta=0.85^{+.2}_{-.1}
\,^{+.3}_{-.1}\,{\rm GeV^{-1}}$ or $g=0.76\,^{+.03}_{-.03} \,^{+.2}_{-.1}$,
\mbox{$\beta=4.90^{+.3}_{-.3} \,^{+5.0}_{-.7}\,{\rm GeV^{-1}}$}. The first
errors are experimental, while the second are estimates of the uncertainty
induced by the counterterms.  (The experimental limit $\Gamma_{D^{*+}} <
0.13\,{\rm MeV}$ excludes the $g=0.76$ solution.) Predictions for the $D^*$ and
$B^*$ widths are given.  

\end{abstract}
}


\newpage 



\section{Introduction}

Combining chiral perturbation theory with heavy quark effective theory (HQET)
gives a good description of the low energy strong interactions between the
pseudo-goldstone bosons and mesons containing a single heavy quark.  Due to
heavy quark symmetry (HQS)\cite{HQS} there is one coupling, $g$, for $D^*D\pi$,
$D^*D^*\pi$, $B^*B\pi$, and $B^*B^*\pi$, and one coupling, $\beta$, for
$D^*D\gamma$, $D^*D^*\gamma$, $B^*B\gamma$, and $B^*B^*\gamma$ at leading
order{\tighten\footnote{Where it is meaningful we use $\pi$ to denote any
member of the pseudo-goldstone boson SU(3) octet, and $D^*$ and $D$ for any
member of the triplets $(D^{*0},D^{*+},D_s^*)$ and $(D^0,D^+,D_s)$ with a
similar notation for $B^*$ and $B$.}}.  The value of the coupling $g$ is
important, since it appears in the expressions for many measurable quantities
at low energy.  These include the rate $B\to D^{(*)}\pi\ell\bar\nu_\ell$
\cite{Bpirate}, form factors for weak transitions between heavy and light
pseudo-scalars \cite{chlag,hlff,FG}, decay constants for the heavy mesons
\cite{dconst,BG}, weak transitions to vector mesons \cite{vector}, form factors
for $B\to D^{(*)}\ell \bar\nu_\ell$ \cite{iwfn}, and heavy meson mass
splittings \cite{msplit} (for a review see \cite{review}).  However, the value
of $g$ has remained somewhat elusive, with numbers in the literature from $\sim
0.2$ to $1.0$.  Recently, a CLEO measurement \cite{CLEO} of $D^{*+}\to
D^+\gamma$ has brought the experimental uncertainties to a level where a model
independent extraction of $g$ is possible from $D^*$ decays.  

As a consequence of HQS the mass splitting between $D^*$ and $D$ mesons is
small (of order $\Lambda_{\rm QCD}^2/m_c$), leaving only a small amount of
phase space for $D^*$ decays. In the dominant modes, $D^*\to D\pi$, and $D^*\to
D\gamma$, the outgoing pion and photon are soft making the chiral expansion a
valid framework.  The branching ratios for $D^{*+}$ decay are
$D^0\,\pi^+$~(67.6\%), $D^+\,\pi^0$~(30.7\%) and $D^+\,\gamma$~(1.7\%)
\cite{CLEO}.  A $D^{*0}$ can only decay into $D^0\,\pi^0$~(61.9\%) and
$D^0\gamma$~(38.1\%) \cite{PDG} since there is not enough phase space for
$D^+\,\pi^-$.  The $D^*_s$ decays predominantly to $D_s\,\gamma$~(94.2\%) with
a small amount going into the isospin violating mode $D_s\,\pi^0$~(5.8\%)
\cite{PDG}.  Since a measurement of the widths of the $D^*$ mesons has not yet
been made, it is only possible to compare the ratios of branching fractions
with theoretical predictions.   The ratio $R_\pi^+={\cal B}(D^{*+}\to
D^0\pi^+)/{\cal B}(D^{*+}\to D^+\pi^0)$ is fixed by isospin to be $R_\pi^+ = 2
|\vec k_{\pi^+}|^3 / |\vec k_{\pi^0}|^3 = 2.199 \pm 0.064$ \cite{CLEO} (where
$\vec k_{\pi^{+,0}}$ are three momenta for the outgoing pions in the $D^*$ rest
frame).  This value is often used in experimental extractions of the branching
ratios to reduce systematic errors.

It is interesting to note that the quark model predictions\cite{qmodel} for
$D^{*0}$ and $D^{*+}$ decays agree qualitatively with the data.  One can
understand, for instance, why the branching ratio ${\cal B}(D^{*+}\to
D^+\gamma)$ is small compared to ${\cal B}(D^{*0}\to D^0\gamma)$.  In the quark
model the photon couples to the meson with a strength proportional to the sum
of the magnetic moments of the two quarks, $\mu_2 = 2/(3\,m_c)-1/(3\,m_d)$
for $D^{*+}\to D^+\gamma$ and $\mu_1 =2/(3\,m_c)+2/(3\,m_u)$ for $D^{*0}\to
D^0\gamma$.  The rate for the former is then suppressed by a factor
\begin{eqnarray}
 \left| \frac{\mu_2}{\mu_1} \right|^2 = (m_u/m_d)^2
   { (m_d/m_c-1/2)^2 \over (m_u/m_c+1)^2 } \simeq 0.04 \:,
\end{eqnarray}
where we have used mass ratios appropriate for constituent quarks, $m_u/m_d
\simeq 1$, $m_d/m_c \simeq m_u/m_c \simeq 1/4$.  This suppression results from
the opposite signs in $\mu_1$ and $\mu_2$, which in turn follow from the
(quark) charge assignments and spin wavefunctions for the heavy mesons.  

In the quark model $g=1$ and $\beta\simeq 3\,{\rm GeV^{-1}}$, while for the
chiral quark model $g=0.75$ \cite{ccmodel}.  Relativistic quark models tend to
give smaller values, $g\sim 0.4$ \cite{rqmodel}, as do QCD sum rules, $g \sim
0.2 - 0.4$ \cite{qcdsr}.

Our purpose here is to use heavy meson chiral perturbation theory at one-loop
to extract the couplings $g$ and $\beta$ from $D^*$ decays. In other words, we
wish to examine how sensitive a model independent extraction of $g$ and $\beta$
is to higher order corrections.  For $D^*\to D\gamma$, analyses beyond leading
order have included the heavy quark's magnetic moment which arises at $1/m_c$
\cite{aetal,cg}, and the leading non-analytic effects from chiral loops
proportional to $\sqrt{m_q}$ \cite{aetal}.  $\sqrt{m_q}$ terms proportional to
both $m_K$ and $m_\pi$ were found to be important.  These effects do not
introduce any new unknown quantities into the calculation of the decay rates. 
For $D^*\to D\gamma$ and the isospin conserving $D^*\to D\pi$ decays the effect
of chiral logarithms, $m_q \ln{(\mu/m_q)}$, have also been considered
\cite{ccsu3}.  These are formally enhanced over other $m_q$ corrections in the
chiral limit, $m_q\to 0$, however, the choice of the scale $\mu$ leads to some
ambiguity in their contribution.  (This scale dependence is cancelled by
unknown couplings which arise at order $m_q$ in the chiral Lagrangian.)  The
isospin violating decay $D_s^*\to D_s\pi^0$ has only been considered at leading
order, where it occurs through $\eta-\pi^0$ mixing\cite{chowise}.  

In this paper the investigation of all $D^*$ decays is extended to one-loop,
including symmetry breaking corrections to order $m_q$ and $1/m_c$.  Further
$1/m_c$ and $m_q$ contributions considered here include the effect of nonzero
$D^*$--$D$ and $D_s$--$D^0$ mass splittings, and the exact kinematics
corresponding to nonzero outgoing pion or photon energy in the loop diagrams. 
(Their inclusion is motivated numerically since $m_{\pi^0} \sim m_{D^*}-m_D
\sim m_{D_s}-m_{D}$, and the decay $D^*\to D\pi^0$ only occurs if $m_{D^*}-m_D
> m_{\pi^0}$.) To simplify the organization of the calculation these splittings
will be included as residual mass terms in our heavy meson propagators. This
gives new non-analytic contributions to the $D^*\to D\pi^0$ and $D^*\to D\gamma$
decay rates.  (To treat the mass splittings as perturbations one can simply
expand these non-analytic functions.) At order $m_q$ there are also analytic
contributions due to new unknown couplings which are discussed.  These new
couplings can, in principle, be fixed using other observables.  We estimate the
effect these unknown couplings have on the extraction of $g$ and $\beta$.  

The calculation of the decay rates to order $m_q$ and $1/m_c$ is taken up in
section~II.  In section~III we compare the theoretical partial rates with the
data to extract the $D^*D\pi$ and $D^*D\gamma$ couplings and discuss the
uncertainty involved.  Predictions for the widths of the $D^{*}$ and $B^{*}$
mesons are also given.  Conclusions can be found in section~IV. 


\section{Decay rates for $D^{*0}$, $D^{*+}$, and $D_{\lowercase{s}}^*$}

In this section we construct the effective chiral Lagrangian that describes the
decays $D^* \to D\pi$ and $D^* \to D\gamma$ to first order in the symmetry
breaking parameters $m_q$ and $1/m_c$.   The eight pseudo-goldstone bosons
$\pi^i$ that arise from the breaking $SU(3)_L \times SU(3)_R \rightarrow
SU(3)_V$ are identified with the pseudoscalar mesons
($\pi^0$,$\pi^+$,$\pi^-$,$K^0$,$\bar K^0$,$K^+$,$K^-$,$\eta$). These can be
encoded in the exponential representation $\Sigma = \xi^2 =\exp(2i \pi^i
\lambda^i/f)$, where $\lambda^i$ are $3\times 3$ matrices such that
\begin{eqnarray}{\tighten
 \pi^i \lambda^i = \left( \begin{array}{ccc}
   \pi^0/\sqrt{2}+\eta/\sqrt{6}  & \pi^+   & K^+ \\
   \pi^-   & -\pi^0/\sqrt{2} +\eta/\sqrt{6}  & K^0 \\
   K^-	   & \bar K^0 & -2 \eta/\sqrt{6} \end{array} \right) \,, } \label{pgb}
\end{eqnarray} 
and $f\sim f_\pi = 130 \,{\rm MeV}$.
For the triplets of heavy mesons ($D^0$, $D^+$, $D_s$) and ($D^{*0}$, $D^{*+}$,
$D_s^*$) we use the velocity dependent fields $P_a(v)$ and $P_a^{*\mu}(v)$\
(a=1,2,3) of HQET.  These are included in a $4\times4$ matrix which transforms
simply under heavy quark symmetry 
\begin{equation}\label{Hdef}
H_a = \frac{1+\vslash}2\, \Big[ P_a^{*\mu} \gamma_\mu 
  - P_a\, \gamma_5 \Big] \,, \label{HQf}
\end{equation} 
and satisfies $\vslash H_a=H_a=-H_a\vslash$.  Including the quark mass term
$m_q={\rm diag}(m_u,m_d,m_s)$ the lowest order Lagrangian is then \cite{chlag}
\begin{eqnarray}
 {\cal L}_0 &=& \frac{f^2}8 {\rm Tr}\, \partial^\mu\Sigma\, \partial_\mu
   \Sigma^\dagger +{f^2 B_0 \over 4}{\rm Tr} (m_q \Sigma + m_q \Sigma^\dagger) 
 -{\rm Tr}\, \bar H_a i v\cdot D_{ba} H_b + 
   g\, {\rm Tr}\, \bar H_a H_b \gamma_\mu \gamma_5 A^\mu_{ba} \,, \label{Lag0}
\end{eqnarray}
where the derivative $D^\mu_{ab}=\delta_{ab}\,\partial^\mu -V_{ab}^\mu$, and
$\bar H_a = \gamma^0 H_a^\dagger \gamma^0$.  The vector and axial vector
currents, $V_{ab}^\mu=\frac12 (\xi^\dagger\partial^\mu\xi + \xi\partial^\mu
\xi^\dagger)$ and $A_{ab}^\mu=\frac{i}2 (\xi^\dagger\partial^\mu\xi -
\xi\partial^\mu \xi^\dagger)$, contain an even and odd number of pion fields
respectively.  The Lagrangian in Eq.~(\ref{Lag0}) is invariant under heavy
quark flavor and spin symmetry. It is also invariant under chiral $SU(3)_L
\times SU(3)_R$ transformations, where $\Sigma\to L\Sigma R^\dagger$, $\xi \to
L \xi U^\dagger = U \xi R^\dagger$, $H\to H U^\dagger$, if we take the quark
mass (which breaks the chiral symmetry) to transform as $m_q \to L m_q
R^\dagger$.  

  The last term in Eq.~(\ref{Lag0}) couples $P^*P\pi$ and $P^*P^*\pi$ with
strength $g$ and  determines the decay rate $D^* \rightarrow D \pi$ at lowest
order.  Going beyond leading order involves including loops with the
pseudo-goldstone bosons, as well as higher order terms in the Lagrangian with
more powers of $m_q$, $1/m_c$, and derivatives.   At order $m_q \sim 1/m_c$ the
following mass correction terms appear
\begin{eqnarray}
 {\cal L}_{m} = \frac{\lambda_2}{4\,m_Q} {\rm Tr}\,\bar H_a 
  \sigma^{\mu\nu} H_a \sigma_{\mu\nu} + 2\lambda_1 {\rm Tr}\,\bar H_a H_b 
  m^\xi_{ba} + 2\lambda_1' {\rm Tr}\,\bar H_a H_a m^\xi_{bb}, \label{Lagm}
\end{eqnarray}
where $m^\xi = \frac12 (\xi m_q \xi^\dagger + \xi^\dagger m_q \xi)$.  The
$\lambda_1'$ term can be absorbed into the definition of $m_H$ by a phase
redefinition of $H$.  The $\lambda_2$ term is responsible for the $D^*$-$D$ mass
splitting at this order, $\Delta=m_{D^{*}}-m_{D}= -2\lambda_2/m_c$.  The term
involving $\lambda_1$ splits the mass of the triplets of $D$ and $D^*$ states. 
Ignoring isospin violation this splitting is characterized by
$\delta=m_{D_s^*}-m_{D^*} = m_{D_s}-m_{D}=2\lambda_1(m_s-\hat m)$ where $\hat
m=m_u=m_d$.  For the purpose of our power counting $\delta \sim m_q \sim 1/m_c
\sim \Delta$.  The effect of these mass splitting terms can be taken into
account by including a residual mass term  in each heavy meson propagator. 
Since we are interested in decay rates we choose the phase redefinition for our
heavy fields to scale out the decaying particle's mass.  For $D^{*0}$ and
$D^{*+}$ decays the denominator of our propagators are: $2v\cdot k$ for
$D^{*0}$ and $D^{*+}$, $2(v\cdot k -\delta)$ for $D_s^*$, $2(v\cdot k+\Delta)$
for $D^0$ and $D^+$, and $2(v\cdot k+\Delta-\delta)$ for $D_s$.  For the
$D_s^*$ decays the denominators are the above factors plus $2\delta$.  (If we
scaled out a different mass then the calculation in the rest frame of the
initial particle would involve a residual `momentum' for the initial particle,
but would yield the same results.)   This results in additional non-analytic
contributions from one-loop diagrams which are functions of the quantities
$\Delta/m_{\pi_i}$ and $\delta/m_{\pi_i}$.  Formally, $m_{\pi_i}^2 \sim m_q
\sim \Delta \sim \delta$ and one can expand these contributions to get back the
result of treating the terms in Eq.~(\ref{Lagm}) as perturbative mass
insertions.  

Another type of $1/m_c$ corrections are those whose coefficients are fixed by
velocity reparameterization invariance \cite{vpi,BG}
\begin{eqnarray}
   \delta{\cal L}_v &=& - \frac1{2m_Q}{\rm Tr}\,\bar H_a (iD)^2_{ba} H_b + 
  \frac{g}{m_Q}{\rm Tr} \bar H_c (i \overleftarrow D^\mu_{ac} v\cdot A_{ba} 
  - i v\cdot A_{ac}\overrightarrow D^\mu_{ba}) H_b \gamma_\mu\gamma_5 \,.
  \label{Lagv}
\end{eqnarray}
The first term here is the HQET kinetic operator, $O_{\rm kin}=\frac1{2m_Q}\bar
h_v\,(iD)^2\, h_v$, written in terms of the interpolating fields $P_a$ and
$P^{*\mu}_a$.  In conjunction with the HQET chromomagnetic operator,
\mbox{$O_{\rm mag}= \frac1{2m_Q}\bar h_v\,\frac{g_s}2\,\sigma_{\alpha\beta}
G^{\alpha\beta}\,h_v$}, these contributions to the Lagrangian modify the
dynamics of the heavy meson states.  They give $1/m_c$ corrections in the form
of time ordered products with the leading order current \cite{cchqs}, which
induce spin and flavor symmetry violating corrections to the form of the
$D^*D\pi$ coupling.  We account for these corrections by introducing the
couplings $g_1$ and $g_2$ in Eq.~(\ref{Lagg}) below.  The last term in
Eq.~(\ref{Lagv}) contributes at higher order in our power counting since it is
suppressed by both a derivative and a power of $1/m_c$.  

Further terms that correct the Lagrangian in Eq.~(\ref{Lag0}) at order $m_q 
\sim 1/m_c$ include \cite{BG}\footnote{The $\kappa_1'$ term was not 
present in \cite{BG}. The factor $B_0/\Lambda_\chi$ is introduced here for 
later convenience.}
\begin{eqnarray}
 \delta{\cal L}_g &=& \frac{g\kappa_1\,B_0}{\Lambda_\chi^2}{\rm Tr}\, 
 \bar H_a H_b \gamma_\mu\gamma_5 A^\mu_{bc} m^\xi_{ca} + \frac{g\kappa_1'\,B_0}
 {\Lambda_\chi^2} {\rm Tr}\, \bar H_a H_b \gamma_\mu\gamma_5 m^\xi_{bc} 
 A^\mu_{ca} \nn\\
 &+& \frac{g\kappa_3\,B_0}{\Lambda_\chi^2}{\rm Tr}\, \bar H_a H_b 
  \gamma_\mu\gamma_5 A^\mu_{ba} m^\xi_{cc} + \frac{g\kappa_5\,B_0}
  {\Lambda_\chi^2}{\rm Tr}\, \bar H_a H_a \gamma_\mu\gamma_5 A^\mu_{bc} 
  m^\xi_{cb} \nn\\ 
 &+& \frac{\delta_2}{\Lambda_\chi}{\rm Tr}\,\bar H_a H_b 
  \gamma_\mu\gamma_5 iv\cdot D_{bc} A_{ca}^\mu + \frac{\delta_3}{\Lambda_\chi}
  {\rm Tr}\,\bar H_a H_b \gamma_\mu\gamma_5 i D^\mu_{bc} v\cdot A_{ca} \nn\\
 &+& \frac{g_1}{m_Q}{\rm Tr}\,\bar H_a H_b \gamma_\mu\gamma_5 A^\mu_{ba}   + 
  \frac{g_2}{m_Q}{\rm Tr}\,\bar H_a \gamma_\mu \gamma_5 H_b A^\mu_{ba}
  + \ldots \,, \label{Lagg}
\end{eqnarray}
where $D^\alpha_{bc}A^\beta_{ca}=\partial^\alpha A^\beta_{ba}+
[V^\alpha,A^\beta]_{ba}$ and $\Lambda_\chi=4\pi f$.  The ellipses here denote
terms linear in $m^{\xi}_-=\frac12 (\xi m_q \xi^\dagger -\xi^\dagger m_q\xi)$
which contribute to processes with more than one pion, as well as terms with
$(iv\cdot D)$ acting on an $H$.  For processes with at most one pion and $H$
on-shell the latter terms can be eliminated at this order, regardless of
their chiral indices, by using the equations of motion for $H$.  The $\kappa_i$
coefficients contain infinite and scale dependent pieces which cancel the
corresponding contributions from the one-loop $D^* \to D\pi$ diagrams.  For the
$\kappa_1$ and $\kappa_1'$ terms only the combination $\tilde\kappa_1 =
\kappa_1+\kappa_1'$ will enter in an isospin conserving manner here. (The
combination $\kappa_1-\kappa_1'$ will contribute an isospin violating
correction to $R^+_\pi$.) At a given scale $\mu$, the finite part of $\kappa_3$
can be absorbed into the definition of $g$.  The decays $D^*\to D\pi$ have
analytic contributions from $\tilde\kappa_1$ and $\kappa_5$ at order $m_q$.

For $m_Q=m_c$ the term in Eq.~(\ref{Lagg}) involving $g_1$ can be absorbed into
$g$ (this term only enters into a comparison with $B^*$ decays).  The term
$g_2$ breaks the equality of the $D^*D\pi$ and $D^*D^*\pi$ couplings.  Since we
only need the coupling $D^*D^*\pi$ in loops we can also absorb $g_2$ into the
definition of $g$.  Thus, our $g$ is defined as the $D^*D\pi$ coupling with
$1/m_Q$ corrections arising in relating it to the couplings for $D^*D^*\pi$ and
$B^{(*)}B^*\pi$.  

The terms in Eq.~(\ref{Lagg}) involving $\delta_2$ and $\delta_3$ contribute
to $D^*\to D\pi^0$, entering in a fixed linear combination with the tree level
coupling $g$ of the form $g - (\delta_2+\delta_3) v\cdot k/ \Lambda_\chi$. 
These are $\sim 10\%$ corrections for the decays $D^* \rightarrow D\pi$.  The
energy of the outgoing pion is roughly the same for all three decays, $v\cdot
k\sim .144 \,{\rm GeV}$.  Therefore, it is impossible to disentangle the
contribution of $\delta_{1,2}$ from that of $g$ for these decays, and the
extraction of $g$ presented here will implicitly include their contribution. 
For other processes involving pions with different $v\cdot k$ these
counterterms can give a different contribution.  This should be kept in mind
when this value of $g$ is used in a different context.

Techniques for one-loop calculations in heavy hadron chiral perturbation theory
are well known and will not be discussed here.  Dimensional regularization is 
used and the renormalized counterterms are defined by subtracting the pole 
terms $1/\epsilon -\gamma+\log{(4\pi)}$.  The decays $D^{*0}\to D^0 \pi^0$ 
and $D^{*+}\to D^+ \pi^0$, and $D_s^* \to D_s \pi^0$ have decay rates
$\Gamma^1_{\pi}$, $\Gamma^2_{\pi}$, and $\Gamma^3_{\pi}$ given by
\begin{eqnarray}
  \Gamma^a_{\pi} &=& {g^2 \over 12\,\pi\,f^2} \left| {Z_{\rm wf}^a \over 
	Z_\pi^{a}} \right|^2 \: |{\vec k}^{\,a}_\pi|^3 \,. \label{ratepi}
\end{eqnarray} 
Here ${\vec k}^{\,a}_\pi$ is the three momentum of the outgoing pion,
$Z_\pi^{a}$ contains the vertex corrections, and $Z_{\rm wf}^a= \sqrt{
Z_{D^*}^a Z_{D}^a }$ contains the wavefunction renormalization for the $D^*$ 
and $D$.  When the ratio of $\Gamma^a_\pi$ to the $D^*\to D\gamma$ rate is
taken $Z_{\rm wf}^a$ will cancel out. However, $Z_{\rm wf}^a$ does contribute
to our predictions for the $D^*$ widths, where the ratio $Z_{\rm
wf}^a/Z_\pi^{a}$ will be kept to order $g^2$. 
\begin{figure}[t!]  
  \centerline{\epsfxsize=15.0truecm \epsfbox{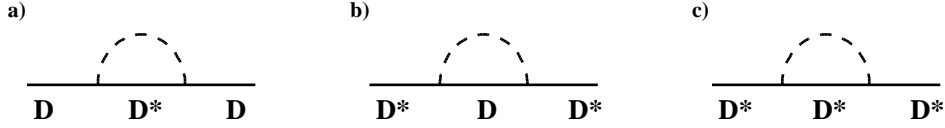}}
{\tighten
  \caption[1]{ $D$ and $D^*$ wavefunction renormalization graphs.  The dashed
	line represents a pseudo-goldstone boson.} \label{fig_wf} }
\end{figure}  
The graphs in Fig.~\ref{fig_wf} give
\begin{eqnarray}
 Z^a_{D} &=& 1 + {g^2 \over (4\pi f)^2} (\lambda^i_{ab} \lambda^{i\dagger}_{ba})
  \bigg\{ [3 m_i^2-6(\Delta+d_0)^2] \log{(\frac{\mu^2}{m_i^2})} +\,
  3\, G_1(m_i,\Delta+d_0) \bigg\} \,, \nn\\ 
 Z^a_{D^*} &=& 1 + {g^2 \over (4\pi f)^2} (\lambda^i_{ab} 
  \lambda^{i\dagger}_{ba}) \bigg\{ [3m_i^2-4d_0^2-2(d_0-\Delta)^2] 
  \log{(\frac{\mu^2}{m_i^2})} + 2\, G_1(m_i,d_0)  \nn\\
 & &\qquad\qquad\qquad\qquad\quad +\, G_1(m_i,d_0-\Delta) 
  \bigg\} \,, \label{wfnren} 
\end{eqnarray}
where $m_i$ is the mass of $\pi^i$, $d_0=\delta^{b3}\delta$ for $D^{*0}$ and
$D^{*+}$ decays and $d_0=(\delta^{b3}-1)\delta$ for $D^*_s$ decays.  The
notation in Eq.~(\ref{wfnren}) assumes that we sum over $b=1,2,3$ and
$i=1,\ldots,8$. The logarithms agree with \cite{ccsu3}, except that we have
kept terms of order $\Delta^2\sim d_0^2$ in the prefactor since these terms are
enhanced for $m_q\to 0$.  Analytic terms of order $\Delta^2 \sim d_0^2$ are
neglected since they are higher order in our power counting.  The function
$G_1(a,b)$ in Eq.~(\ref{wfnren}) has mass dimension $2$. It contains an
analytic part proportional to $a^2$, and a non-analytic part which is a function
of the ratio $b/a$.    The expression for $G_{1}$ can be found in the Appendix.

\begin{figure}[t]  
  \centerline{\epsfxsize=18.0truecm \epsfbox{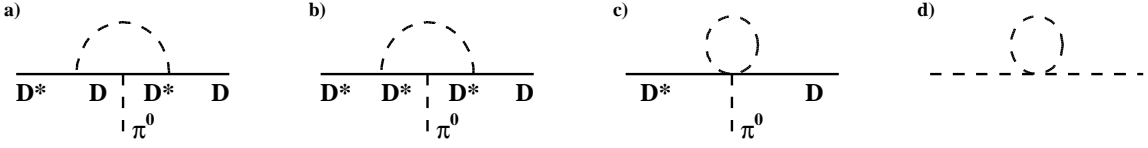}}
{\tighten
  \caption[1]{Nonzero one-loop vertex corrections for the decays $D^{*0}\to 
     D^0\pi^0$ and $D^{*+}\to D^+\pi^0$ (a,b,c) and the pseudo-goldstone boson 
     wave function renormalization graph (d). } \label{fig_pi12} }
\end{figure}

For $a=1,2$ the decay proceeds directly so that at tree level $Z_{\rm wf}^{1,2}
/Z_\pi^{1,2}=1$.  At one loop we have non-zero vertex corrections from the
graphs in Fig.~\ref{fig_pi12}a,b,c.  As noted in \cite{ccsu3}, the two one-loop
graphs that contain a $D^{(*)}D^*\pi\pi$ vertex (not shown) vanish, and the
graph in Fig.~\ref{fig_pi12}c cancels with the $\pi^0$ wavefunction
renormalization in Fig.~\ref{fig_pi12}d (this is also true for $D^{*+}\to
D^0\pi^+$ and $D_s^*\to D_s\pi^0$).  Therefore for $a=1,2$ the vertex 
corrections are
\begin{eqnarray}
 {1\over Z_\pi^{\,a}} &=& 1 + {g^2 \over (4\pi f)^2} {\lambda^i_{ab}
  \lambda^1_{bb} \lambda^{i\dagger}_{ba} \over \lambda^1_{aa}} \Bigg\{ 
  \log{(\frac{\mu^2}{m_i^2})} \Big[ m_i^2 +\frac23(-d_1^2+d_1\,d_2+d_2^2
  -2 d_1\,d_0-2\,d_0^2) \Big] \nn \\
 & &+ 2\,F_1(m_i,d_1,d_2)-4\,F_1(m_i,d_1,d_0) \Bigg\} 
  + \varrho_{\pi}^a(\tilde\kappa_1,\kappa_5) \,, \label{pi12vertex}
\end{eqnarray}
where here $d_0=\delta^{b3}\delta$, $d_1=k\cdot v + d_0$, $d_2=-\Delta+d_0$,
and $k$ is the outgoing momentum of the $\pi^0$.  The coefficient of the
$m_i^2\log{(\mu^2/m_i^2)}$ term agrees with \cite{ccsu3}.  The function $F_1$
has mass dimension $2$ and contains both analytic and non-analytic parts. 
$\varrho_{\pi\,ct}^a$ contains the dependence of the rate on the (renormalized)
counterterms $\tilde\kappa_1(\mu)$ and $\kappa_5(\mu)$.   With isospin
conserved $\varrho_{\pi}^{1,2}$ do not depend on $\kappa_5$, and furthermore
are proportional to $m_\pi^2/(4\pi f)^2$, so these counterterms are small.
Expressions for $F_1$ and $\varrho_\pi^a$ are given in the Appendix.  

\begin{figure}[tb]  
  \centerline{\epsfxsize=18.0truecm \epsfbox{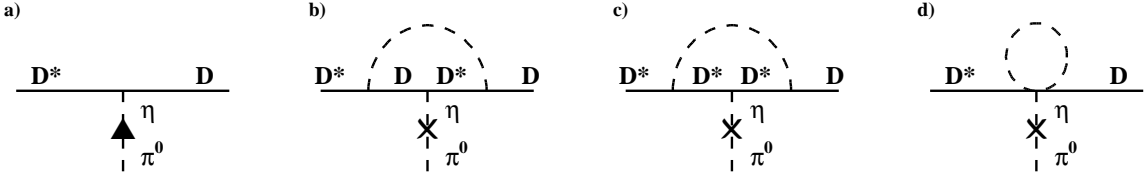}}
{\tighten
  \caption[1]{Nonzero vertex corrections  for the decay $D_s^*\to D_s\pi^0$ 
     which involve  $\pi^0-\eta$ mixing. The cross denotes leading order 
      mixing while the triangle denotes mixing at next to leading order.} 
 \label{fig_pi3} }
\end{figure}

The decay $D_s^*\to D_s\pi^0$ is isospin violating, and the leading
contribution occurs through $\eta-\pi^0$ mixing\cite{chowise}.  To first order
in the isospin violation the decay is suppressed at tree level by the mixing 
angle $\theta = (1.00\pm 0.05)\times 10^{-2} $ \cite{GLrev}
\begin{eqnarray}
 {1\over Z_\pi^{3}}= {(m_u-m_d)\over 2\,(m_s-\hat m)} = -\frac{2}{\sqrt{3}}
  \theta \simeq -\frac1{87.0} \,. \label{pi3vertex0}
\end{eqnarray}
Beyond tree level we have corrections to the $\eta-\pi^0$ mixing angle
parameterized by $\delta_{mix}=0.11$ \cite{GL} (Fig.~\ref{fig_pi3}a), loop
corrections to the $\eta-\pi^0$ mixing graph (Figs.~\ref{fig_pi3}b,c,d), as
well as loop graphs with decay directly to $\pi^0$ that occur in an isospin
violating combination (Figs.~\ref{fig_pi12}a,b).  The contribution of
Fig.~\ref{fig_pi3}d is again cancelled by the pseudo-goldstone boson wave
function renormalization graph (Fig.~\ref{fig_pi12}d). 
Note that the decay $D_s^*\to D_s\pi^0$ cannot occur via a single virtual 
photon in the effective theory.  In the quark model, decay to the spin and
color singlet $\pi^0$ can occur if the single photon is accompanied by at least
two gluons (with suppression $\alpha/\pi\simeq 1/430$ \cite{chowise}).  We will
neglect the possibility of such a single photon mediated transition here. Thus,
\begin{eqnarray}
 {1\over Z_\pi^{\,3}} &=& {(m_u-m_d) \over 2\,(m_s-\hat m)} \Bigg[ 1 + 
  \delta_{mix} + {g^2 \over (4\pi f)^2} { \lambda^i_{3b}
  \lambda^8_{bb} \lambda^{i\dagger}_{b3} \over \lambda^8_{33} } \Bigg\{  
  \log{(\frac{\mu^2}{m_i^2})} \Big[ m_i^2 +\frac23(-d_1^2+d_1\,d_2+d_2^2 \nn\\
 & & \qquad\qquad\qquad -2 d_1\,d_0-2\,d_0^2) \Big] + 
  2\,F_1(m_i,d_1,d_2)-4\,F_1(m_i,d_1,d_0) \Bigg\} \Bigg] \\ \label{pi3vertex}
 & &+{g^2 \over (4\pi f)^2} { \lambda^i_{3b} \lambda^1_{bb} 
  \lambda^{i\dagger}_{b3} \over (1/\sqrt{2})} \Bigg[ \tilde m_i^2
  \log{(\frac{\mu^2}{\tilde m_i^2})} + 2\,F_1({\tilde m}_i,d_1,d_2)-4\,
  F_1({\tilde m}_i,d_1,d_0) \Bigg] \nn\\
 & & + \varrho_{\pi}^3(\tilde\kappa_1,\kappa_5) \nn \,,  
\end{eqnarray}
where for $D_s^*$ decay $d_0=(\delta^{b3}-1)\delta$, $d_1=k\cdot v + d_0$, and
$d_2=-\Delta+d_0$.  The tilde on the mass, $\tilde m_i$, indicates that isospin
violation is taken into account.  Note that $\sqrt{2}\:\sum_{i,b}\lambda^i_{3b}
\lambda^1_{bb} \lambda^{i\dagger}_{b3} \tilde m_i^2 = m_{K^\pm}^2-m_{K^0}^2$. 
The function $\varrho_\pi^3$ depends on $\tilde\kappa_1$, $\kappa_5$, 
and has both $m_K^2$ and $m_\pi^2$ terms.

 To describe $D^* \to D\gamma$, electromagnetic effects must be
included, so the Lagrangian in Eq.~(\ref{Lag0}) is gauged with a $U(1)$
photon field $B^\mu$.  With octet and singlet charges, $Q={\rm
diag}(\frac23,-\frac13,-\frac13)$ and $Q'= \frac23$ (for the $c$), the
covariant derivative ${\cal D_\mu}$ is defined as \cite{ccem} ${\cal D_\mu}\xi =
\partial_\mu\xi + ieB_\mu[Q,\xi]$ and ${\cal D_\mu} H = \partial_\mu H +ieB_\mu
(Q'H - H Q) - {\cal V_\mu}H$, where the vector and axial vector currents are
now ${\cal V_\mu}= \frac12 (\xi^\dagger {\cal D_\mu}\xi + \xi{\cal D_\mu}
\xi^\dagger)$ and ${\cal A_\mu}=\frac{i}2 (\xi^\dagger{\cal D_\mu}\xi -
\xi{\cal D_\mu} \xi^\dagger)$.  However, this procedure does not induce a
coupling between $D^*$, $D$ and $B_\mu$ without additional pions.  Gauge
invariant contact terms should also be included, and it is one of these that
gives rise to the $D^*D\gamma$ coupling (and a $D^*D^*\gamma$ coupling)
\begin{eqnarray}
 {\cal L}_\beta = \frac{\beta\,e}{4}{\rm Tr}\,\bar H_a H_b 
  \sigma^{\mu\nu} F_{\mu\nu} Q^\xi_{ba} . \label{Lagb}
\end{eqnarray}
Here $\beta$ has mass dimension $-1$, $Q^\xi = \frac12 (\xi^\dagger Q \xi
+ \xi Q \xi^\dagger)$, and $F_{\mu\nu}=\partial_\mu B_\nu-\partial_\nu B_\mu$. 
The terms which correct this Lagrangian at order $m_q \sim 1/m_c$ have a 
similar form to those in Eq.~(\ref{Lagg})
\begin{eqnarray}
 \delta{\cal L}_\beta &=&\frac{\alpha_1\,B_0}{\Lambda_\chi^2}\frac{\beta\,e}{4}
  {\rm Tr}\, \bar H_a H_b \sigma^{\mu\nu}F_{\mu\nu} Q^\xi_{bc} m^\xi_{ca} 
  + \frac{\alpha_1'\,B_0}{\Lambda_\chi^2} \frac{\beta\,e}{4} {\rm Tr}\, 
  \bar H_a H_b \sigma^{\mu\nu}F_{\mu\nu}  m^\xi_{bc} Q^\xi_{ca} \nn\\
 &+& \frac{\alpha_3\,B_0}{\Lambda_\chi^2}\frac{\beta\,e}{4}{\rm Tr}\, 
  \bar H_a H_b \sigma^{\mu\nu} F_{\mu\nu} Q^\xi_{ba} m^\xi_{cc} 
  + \frac{\alpha_5\,B_0}{\Lambda_\chi^2}\frac{\beta\,e}{4}{\rm Tr}\,
  \bar H_a H_a \sigma^{\mu\nu} F_{\mu\nu} Q^\xi_{bc} m^\xi_{cb} \nn\\
 &+& \frac{\tau_2\,e}
  {4\Lambda_\chi^2}{\rm Tr}\,\bar H_a H_b \sigma^{\mu\nu} Q^\xi_{bc}
  iv\cdot D_{ca} F_{\mu\nu}
  + \frac{\tau_3\,e}{4\Lambda_\chi^2}
  {\rm Tr}\,\bar H_a H_b \sigma^{\mu\nu}  Q^\xi_{bc} i D^\mu_{ca} v^\lambda 
  F_{\nu\lambda} \nn\\
 &+& \frac{\beta_1}{m_Q}\frac{e}{4}{\rm Tr}\,\bar H_a H_b 
  \sigma^{\mu\nu} F_{\mu\nu} Q^\xi_{ba}
  + \frac{\beta_2}{m_Q}\frac{e}{4}{\rm Tr}\,
  \bar H_a \sigma^{\mu\nu} H_b  F_{\mu\nu} Q^\xi_{ba} \nn\\
 &-& \frac{e}{4\,m_Q} Q'\, 
  {\rm Tr}\, \bar H_a \sigma^{\mu\nu} H_a F_{\mu\nu} + \ldots\,. \label{Lagdb} 
\end{eqnarray} 
The ellipses denote terms that do not contribute for processes without
additional pions and/or can be eliminated using the equations of motion for
$H$.  For our purposes $Q^\xi$ and $m^\xi$ in Eq.~(\ref{Lagdb}) are diagonal so
only $\tilde \alpha=\alpha_1+\alpha_1'$ contributes.  The finite part of
$\alpha_3$ will be absorbed into the definition of $\beta$.  For $m_Q=m_c$, the
$\beta_1$ term can be absorbed, and we absorb the part of the $\beta_2$ term
that contributes to $D^*D\gamma$ since $D^*D^*\gamma$ only contributes in loops
for us.  Thus, $\beta$ is defined to be the $D^*D\gamma$ coupling at order
$1/m_c$.  The last term in Eq.~(\ref{Lagdb}) is the contribution from the
photon coupling to the $c$ quark and has a coefficient which is fixed by heavy
quark symmetry \cite{HQS}.  The $\tau_{1,2}$ terms are similar to the
$\delta_{2,3}$ terms in Eq.~(\ref{Lagg}), and appear with $\beta$ in the
combination $\beta - (\tau_1+\tau_2) v\cdot k/ \Lambda_\chi^2$.  Here
$\tau_1+\tau_2$ will have an infinite part necessary for the one-loop
renormalization.  Again it is not possible to isolate the finite part of the
$(\tau_1+\tau_2)$ contribution from that of $\beta$, so the extraction at this
order includes the renormalized $\tau_{1,2}$ with $v\cdot k \sim 0.137\,{\rm
GeV}$. 

\begin{figure}[t!]  
  \centerline{\epsfxsize=18.0truecm \epsfbox{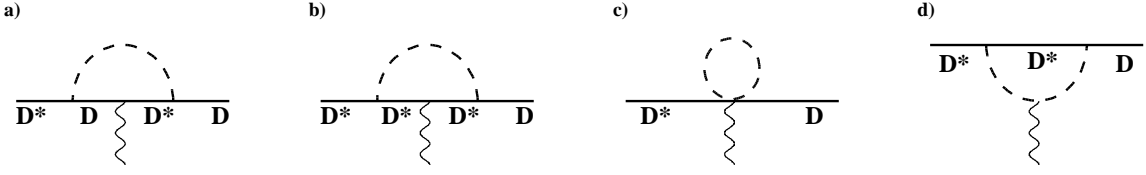}}
 {\tighten
 \caption[1]{Nonzero vertex corrections for the decays $D^*\to D\gamma$. }  
  \label{diag_gam} }
\end{figure}

The decays $D^{*0}\to D^0\gamma$, $D^{*+}\to D^+\gamma$, and $D_s^*\to 
D_s\gamma$ have decay rates $\Gamma_\gamma^1$, $\Gamma_\gamma^2$, and 
$\Gamma_\gamma^3$ given by
\begin{eqnarray}
 \Gamma^a_\gamma &=& \frac{\alpha}3\: |\mu_a|^2\: |{\vec k}^{\,a}_\gamma|^3, 
  \qquad \qquad
  \mu_a = Z_{\rm wf}^a\left( \beta \,{ Q_{aa} \over Z_\gamma^a } +{Q' \over m_c}
  \right) \,, \label{rateg}
\end{eqnarray}
where $\alpha \simeq 1/137$, ${\vec k}^{\,a}_\gamma$ is the three momentum of
the outgoing photon, and the wavefunction renormalization, $Z_{\rm wf}^a$, is
given by Eq.~(\ref{wfnren}).  To predict the $D^*$ widths, $Z_{\rm
wf}^a/Z_\gamma^a$ is kept to order $g^2$ and we take $Z_{\rm wf}^a\times 1/m_c
= 1/m_c$. The vertex correction factor $Z_\gamma^a$ has nonzero contributions
from the graphs in Fig.~\ref{diag_gam}.  Note that the two one-loop graphs that
contain a $D^{(*)}D^*\pi\gamma$ vertex (not shown) do not contribute
\cite{ccsu3}.  
Furthermore, the graph in Fig.~\ref{diag_gam}c has no contribution from the
$D^*D^*\gamma$ coupling which arises from gauging the lowest order Lagrangian
in Eq.~(\ref{Lag0}).  Thus
\begin{eqnarray}
 {1\over Z_\gamma^{\,a}} &=& 1 + {g^2 \over (4\pi f)^2} {\lambda^i_{ab} Q_{bb} 
  \lambda^{i\dagger}_{ba} \over Q_{aa}} \Bigg\{ 
  \log{(\frac{\mu^2}{m_i^2})} \Big[ m_i^2 +\frac23(-d_1^2+d_1\,d_2+d_2^2
  -2 d_1\,d_0-2\,d_0^2) \Big] \nn \\*
 & &+ 2\,F_1(m_i,d_1,d_2)-4\,F_1(m_i,d_1,d_0) \Bigg\} - {1\over 
  (4\pi f)^2} {[\lambda^{i\dagger},[Q,\lambda^i]]_{aa} \over 2\,Q_{aa}} 
  \bigg[m_i^2\,\log{(\frac{\mu^2}{m_i^2})+m_i^2 \bigg] } \nn\\*
 & &+ {4\,g^2 \over (4\pi f)^2} {(\lambda^i_{ab}\lambda^{i\dagger}_{ba}) 
  \,q^i \over \beta Q_{aa}} \Big[ -
  \log{(\frac{\mu^2}{m_i^2})}(d_0+\frac{k\cdot v}2) + 
  F_2(m_i,d_0,k\cdot v) \Big] \nn\\*
 & & +  \varrho_{\gamma}^a(\tilde\alpha_1,\alpha_5) \,, \label{vertexg}
\end{eqnarray}
where $q^i$ is the charge of meson $\pi^i$, $k$ is now the outgoing photon
momentum, and the $d_i$ are as above (again they differ depending on whether it
is $D^*_s$ or one of $D^{*0}$, $D^{*+}$ that is decaying).  The coefficients of
the $m_i^2\log{(\mu^2/m_i^2)}$ terms agree with \cite{ccsu3}. The new function
$F_2$ has mass dimension $1$.  It contains an analytic part proportional to
$2d_0+v\cdot k$, and a non-analytic part which is a function of $\delta/m_i$
and $v\cdot k/m_i$.  $\varrho^a_\gamma$ contains the dependence of the rate
on the (renormalized) counterterms $\tilde\alpha_1(\mu)$ and $\alpha_5(\mu)$. 
Expressions for $F_2$ and $\varrho_\gamma^a$ are given in the Appendix.

By examining Eqs.~(\ref{wfnren}), (\ref{pi12vertex}), (\ref{pi3vertex}), and
(\ref{vertexg}) we can get an idea of the size of the various one-loop
corrections to $\Gamma_\pi^a$ and $\Gamma_\gamma^a$.  With our power
counting $\Delta \sim \delta \sim v\cdot k \sim m_q \sim m_i^2$ so we can 
consider expanding in $\Delta/m_i$, $\delta/m_i$, and $v\cdot k/m_i$.  Using 
the expressions from the Appendix gives
\begin{eqnarray}
  G_1(m_i,b) &=& \frac{m_i^2}3 \left[ 1- \frac{6\pi b}{m_i} + 
    \frac{16 b^2}{m_i^2} + \ldots \right] \:\,, \nn\\
  F_1(m_i,b,c) &=& -\frac{m_i^2}2 \left[ 1 - \frac{\pi(b+c)}{m_i} +
    \frac{16 (b^2+b c+c^2)}{9 m_i^2} +\ldots \right]  \:\,,   \nn\\
  F_2(m_i,d_0,k\cdot v) &=& -\pi m_i \left[ 1 -\frac{3d_0^2+3 d_0 k\cdot v 
    +(k\cdot v)\,^2}{6 m_i^2}+\ldots 
    \right] \:\,.  \label{expns}
\end{eqnarray}
The leading terms in $G_1$ and $F_1$ are $m_q$ corrections to the rates.  The
second terms are order $m_q^{3/2}$ and $\sqrt{m_q}/m_c$, and can be kept since
they are unambiguously determined at the order we are working.  The third and
remaining terms in $G_1$ and $F_1$ are subleading in our power counting.  The
term $-\pi m_i$ in $F_2$ is the formally enhanced contribution discovered in
\cite{aetal}.  Note that there are no contributions to $F_2$ proportional to
$\delta$ or $k\cdot v$.  The second term in $F_2$ in Eq.~(\ref{expns}) has
contributions of order $m_q^{3/2}$, $\sqrt{m_q} k\cdot v$, and $(k\cdot
v)^2/\sqrt{m_q}$ which again can be kept since they are unambiguously 
determined.

The above power counting is sensible when $m_i$ is $m_K$ or $m_\eta$.  We know
that numerically $m_{\pi} \sim \Delta \sim \delta \sim k\cdot v$, so for
$m_i=m_\pi$ the series in Eq.~(\ref{expns}) are not sensible.  In \cite{aetal}
the term $-\pi m_\pi$ in $F_2$ was found to be important, so we want to keep
corrections with $m_\pi$ dependence.  Therefore, instead of expanding the
non-analytic functions we choose to keep them in the non-analytic forms given in
the Appendix. Numerically the one-loop corrections to $\Gamma_\pi^1$ and
$\Gamma_\pi^2$ are very small; with $g=1$ they are of order $\sim 2\%$.  For
$\Gamma_\pi^3$, $\delta_{mix}$ is a $11\%$ correction to the tree level result
in Eq.~(\ref{pi3vertex0}).  Individually the terms proportional to $g^2 F_1$
and $g^2 \log{(\mu/m_q)}$ in Eq.~(\ref{pi3vertex}) are $\sim 10\%$ corrections
for $g=1$.  However, the loops graphs with $\eta-\pi^0$ mixing tend to cancel
those without $\eta-\pi^0$ mixing leaving a $\sim 2\%$ correction.  The
one-loop corrections to $\Gamma_\gamma^a$ are larger, for instance the graph in
Fig.~\ref{diag_gam}c gives sizeable corrections that are not suppressed by
$g^2$.  Corrections to the coefficient of the leading $g^2/\beta$ term obtained
in \cite{aetal} range from $\sim 3\%$ for $D^*_s$ and $\sim 20\%$ for $D^{*0}$
decay, to $\sim 50\%$ for the $D^{*+}$.  (The latter percentage is large
because the only contribution for this decay come from a charged pion in the
loop of Fig.~\ref{diag_gam}d.)  Corrections proportional to $g^2$ are only
sizeable for $D^*_s\to D_s\gamma$ where they are $\sim 10\%$ for $g=1$.

\section{Extraction of $\lowercase{g}$ and $\beta$}

Using the calculation of the decay rates from the previous section, the
couplings $g$ and $\beta$ can be extracted from a fit to the experimental data.
Input parameters include $m_c=1.4\,{\rm GeV}$ \cite{glkw}, the meson masses
from \cite{PDG}, $\Delta=m_{D^*}-m_D=0.142\,{\rm GeV}$,
$\delta=m_{D_s^{(*)}}-m_{D^{(*)}}=0.100\,{\rm GeV}$, and $v\cdot k$ which is
determined from the masses.  When isospin is assumed we use
$m_K=0.4957\,{\rm GeV}$ and $m_\pi=0.1373\,{\rm GeV}$.  $f$ is extracted from
$\pi^-$ decays. At tree level we use $f=f_\pi=0.131\,{\rm GeV}$ \cite{PDG},
while when loop contributions are included we use the one-loop relation between
$f$ and $f_\pi$ \cite{GL} to get $f=0.120\,{\rm GeV}$.  The ratio of the decay
rates $\Gamma_\gamma^a$ and $\Gamma_\pi^a$ are fit to the experimental numbers
\begin{eqnarray}
 {{\cal B}(D^{*0}\to D^0\gamma) /{\cal B}(D^{*0}\to D^0\pi^0)} &=&
   0.616 \pm 0.076 \ \cite{PDG} \,, \nn\\
 {{\cal B}(D^{*+}\to D^+\gamma) / {\cal B}(D^{*+}\to D^+\pi^0)} &=&
   0.055 \pm 0.017 \ \cite{CLEO} \,, \nn\\
 {{\cal B}(D^{*\ }_s\to D_s\pi^0) / {\cal B}(D^{*\ }_s\to D_s\gamma)} &=&
   0.062 \pm 0.029 \ \cite{PDG} \,, \label{exptratios}
\end{eqnarray}
where the errors combine both statistical and systematic. Using the 
masses $m_{D^{*0}}$, $m_{D^{*+}}$, $m_{D^{*}_s}$, and mass splittings
$m_{D^{*0}}-m_{D^0}$, $m_{D^{*+}}-m_{D^+}$, $m_{D^{*}_s}-m_{D_s}$ from
\cite{PDG} gives the momentum ratios that appear in 
$\Gamma_\gamma^a/\Gamma_\pi^a$:
\begin{eqnarray}
  {|\vec k_\gamma^1|^3\over |\vec k_\pi^1|^3} = 32.65 \pm 0.44, \qquad  
  {|\vec k_\gamma^2|^3\over |\vec k_\pi^2|^3} = 45.2 \pm 1.0, \qquad  
  {|\vec k_\gamma^3|^3\over |\vec k_\pi^3|^3} = 24.4 \pm 1.5 \,. \label{exptmom}
\end{eqnarray}
The errors here are clearly dominated by those in Eq.~(\ref{exptratios}).
Equating the numbers in Eq.~(\ref{exptratios}) to the ratio of rates from
Eqs.~(\ref{ratepi}) and (\ref{rateg}) gives a set of three nonlinear equations
for $g$ and $\beta$ (where we ignore for the moment the unknown counterterms). 
In general any pair of these equations will have several possible solutions. 
To find the best solution we take the error from Eq.~(\ref{exptratios}) and
minimize the $\chi^2$ for the fit to the three measurements. We will restrict
ourself to the interesting range of values, $0 < g < 1$ and $0 < \beta < 6$,
discarding any solutions that lie outside this range. (The sign of $g$ will not
be determined here since it only appears quadratically in $\Gamma_\pi^a$ and
$\Gamma_\gamma^a$.)

{\tighten
\begin{table}[!t]
\begin{center}
\begin{tabular}{clcccccccccc}  
 \multicolumn{2}{c}{Order} && $g$ & $\beta({\rm GeV^{-1}})$ & $\chi^2$ &&&
      $g$ &  $\beta({\rm GeV^{-1}})$ & $\chi^2$ &  \\ \hline 
 & tree level &&  
  \multicolumn{2}{c}{  $\beta/g=3.6$} & $30.$ && &   &  &  \\
 & +$Q'/m_c$ + one-loop with $\sqrt{m_q}$ && $0.23$ & $0.89$ & $4.3$
   &&& $0.45$ & $2.8$ & $3.7$ & \\
 & + chiral logs && $0.25$ & $0.78$ & $4.1$ &&& $0.56$ & $3.2$ & $1.4$ & \\
 & one-loop with nonzero $\Delta$,$\delta$,$v\cdot k$, 
   && $0.25$ & $0.86$ & $3.9$ &&& $0.83$ & $6.0$ & $2.5$ & \\[-2pt]
 & \quad without analytic $m_q$ terms \\
 & order $m_q\sim 1/m_c$ with  && $0.265$ & $0.85$ & $3.0$ &&& 
   $0.756$ & $4.9$ & $3.9$ & \\
 & \quad $\tilde\kappa_1=\kappa_5=\tilde\alpha_1=\alpha_5=0$
\end{tabular}
\end{center}
{\tighten
\caption{Solutions for $g$ and $\beta$ which minimize the $\chi^2$ associated
 with a fit to the three ratios in Eq.~(\ref{exptratios}). There are two 
 solutions in the region of interest.}
\label{table_solns} }
\end{table} }
To test the consistency of the chiral expansion we will first check how the
extraction of $g$ and $\beta$ differs at various orders.  The results are given
in Table~\ref{table_solns}.  At tree level only the ratio $\beta/g$ is
determined, and the $\chi^2$ is rather large.  We might next consider adding
the contribution from the chiral loop corrections to $D^*\to D\gamma$ which go
as $\sqrt{m_q}$.  However, this does not lead to a consistent solution between
the three data points unless $\beta$ is negative.  This signals the importance
of the $Q'/m_c$ contribution in Eq.~(\ref{rateg}) corresponding to a nonzero
heavy quark magnetic moment.  Adding this contribution gives the results in the
second row of Table~\ref{table_solns}, where there are now two solutions with
similar $\chi^2$ in the region of interest.  Adding the chiral logarithms,
$m_q\log{(\mu/m_q)}$, at scale $\mu=1\,{\rm GeV}$ gives the solutions in the
third row.  Taking nonzero $\delta$, $\Delta$, and $v\cdot k$ in the
non-analytic functions $F_1$ and $F_2$ gives the solutions in the fourth row of
Table~\ref{table_solns}, where the value of $g$ in the second solution has
increased by $\sim 50\%$.  For these two solutions only the analytic $m_i^2$
dependence has been neglected.  Finally, the solutions in row five include the
analytic $m_i^2$ dependence with the counterterms set to zero (at $\mu=1\,{\rm
GeV}$).  The uncertainty associated with these counterterms will be
investigated below.   It is interesting to note that the extracted value of $g$
in the second column of Table~\ref{table_solns} changes very little with the
addition of the various corrections.

\begin{figure}[!t]  
  \centerline{\epsfxsize=9.0truecm \epsfbox{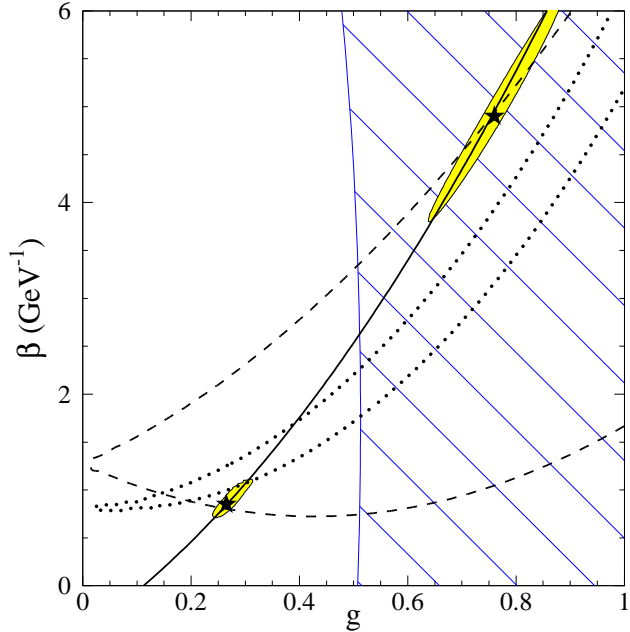} }
 {\tighten
 \caption[1]{Solution contours in the $g$-$\beta$ plane for the situation in
row 5 of Table~\ref{table_solns}. The solid, dashed, and dotted lines
correspond to solution lines for the $D^{*0}$, $D^{*+}$, and $D^*_s$ decay rate
ratios respectively.  The stars correspond to the minimal $\chi^2$ solutions 
and the shaded regions correspond to the 68\% confidence level 
of experimental error in the fit.  The hatched region is excluded by the 
experimental limit $\Gamma(D^{*+}) < 0.13\,{\rm MeV}$ \cite{ACCMOR}.} 
\label{fig_solns} }
\end{figure}
\begin{figure}[!t]  
  \centerline{\epsfxsize=8.0truecm \epsfbox{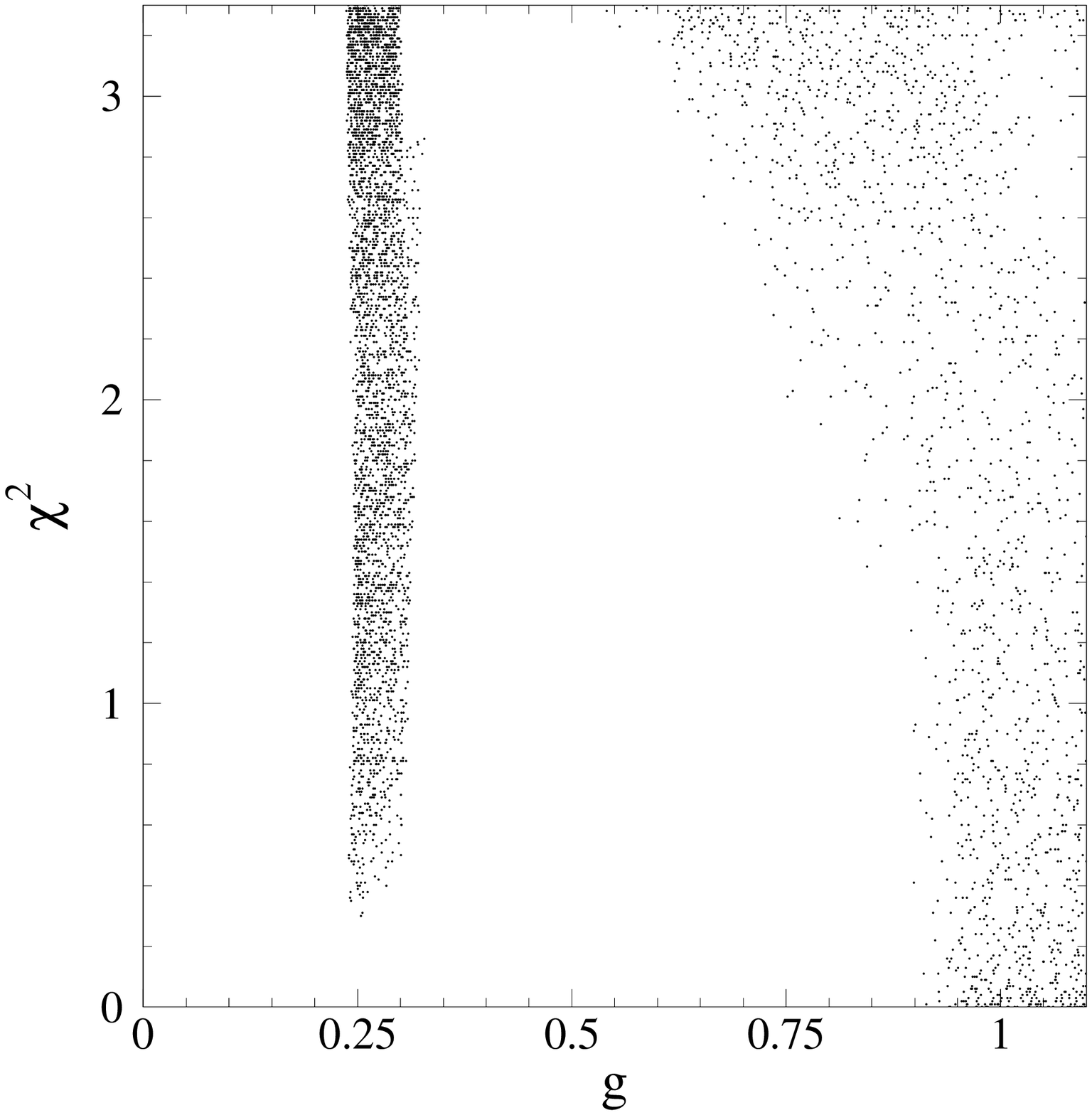}
	\epsfxsize=8truecm \epsfbox{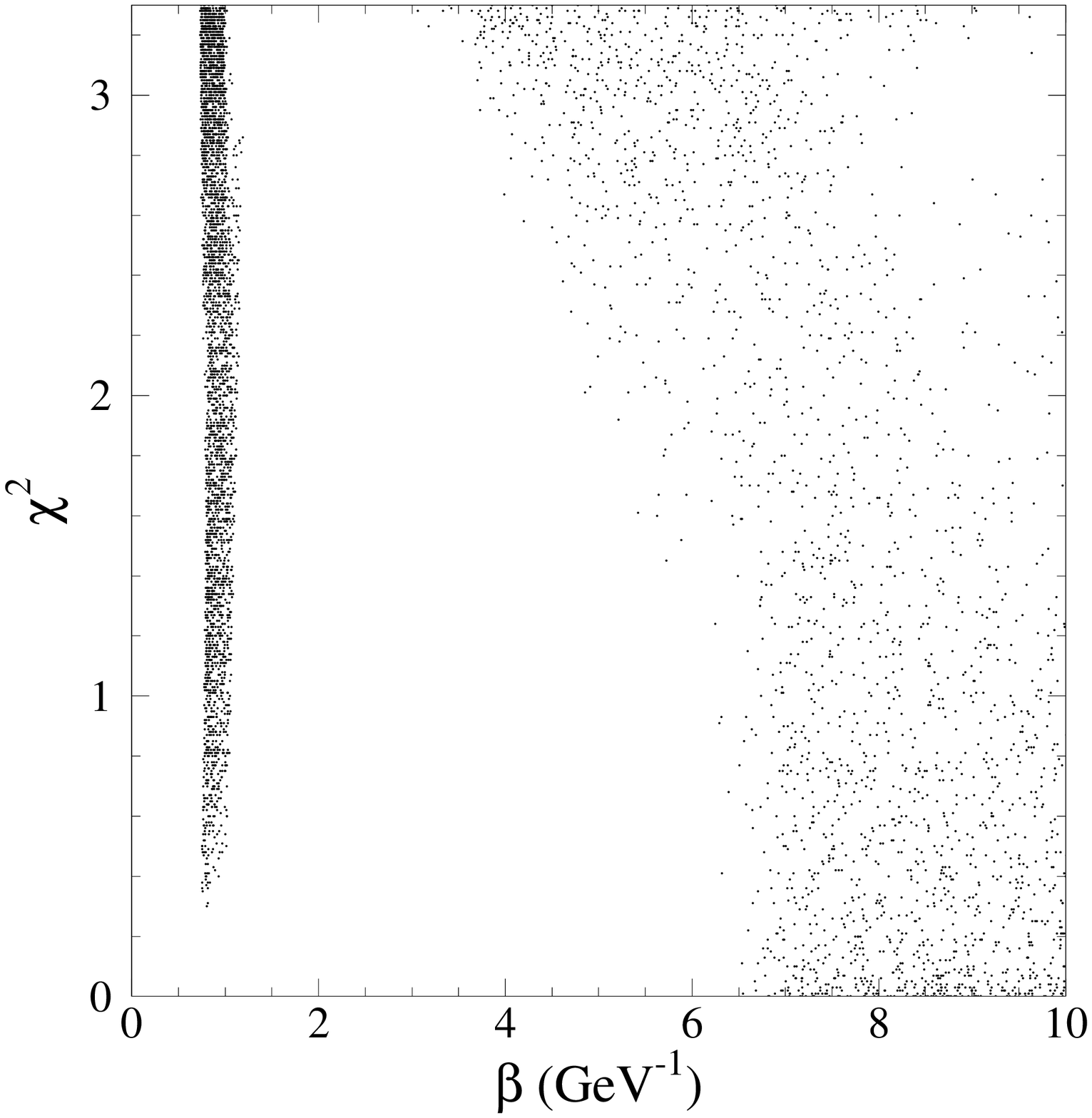} }
 {\tighten  
\caption[1]{Effect of the order $m_q$ counterterms ($\tilde\kappa_1$,
$\kappa_5$, $\tilde\alpha_1$, and $\alpha_5$) on the solutions in
Eq.~(\ref{soln1}).  The counterterms are taken to be randomly distributed with
$-1 < \tilde\kappa_1,\kappa_5 < 1$, $-2 < \alpha_1,\alpha_5 < 2$.  For each set
of counterterms $g$ and $\beta$ were determined at the new minimal $\chi^2$.
$5000$ sets were generated near each of the two solutions.} \label{fig_rand} }
\end{figure}

One can see more clearly how these solutions are determined by looking at
Fig.~\ref{fig_solns}.  The central value for each ratio of decay rates in
Eq.~(\ref{exptratios}) gives a possible contour in the $g$-$\beta$ plane, as
shown by the solid ($D^{*0}$), dashed ($D^{*+}$), and dotted ($D_s^*$) lines. 
An exact solution for two of the ratios occurs at the intersection of two of
these contour lines.  However, a good solution for all three ratios requires a
point that is close to all three lines.  The solutions in the fifth row of
Table~\ref{table_solns} are indicated by stars in Fig.~\ref{fig_solns}.    The
size of the experimental uncertainties can be seen in the 68\% confidence level
ellipses which are shown as shaded regions in the figure (for two degrees of
freedom they correspond to $\chi^2 \le \chi^2_{min}+2.3$).  These regions are
centered on the solid line since the $D^{*0}$ ratio has the smallest
experimental error.  The errors in Eq.~(\ref{exptratios}) give the following
one sigma errors on the two solutions
\begin{eqnarray}
 g=0.265^{+.036}_{-.018} \ \ \ \beta=0.85^{+.21}_{-.10}\,{\rm GeV^{-1}} \:,
    \qquad\quad
 g=0.756^{+0.028}_{-0.027} \ \ \ \beta=4.90^{+.27}_{-.26}\,{\rm GeV^{-1}} \:.
  \label{soln1}
\end{eqnarray}
Both solutions fit the first two ratios in Eq.~(\ref{exptratios}), but do not
do as well for the third.  Minimizing the $\chi^2$ has biased against the third
ratio as a result of its large experimental error.  For this ratio the
$g=0.265$ and $g=0.76$ solutions give values which are $4$ and $13$ times too
small respectively.  For the first solution it is possible to improve the fit
to the third ratio with reasonably sized counterterms.  For instance, simply
taking $\tilde\alpha_1=2$ gives ${{\cal B}(D^{*\ }_s\to D_s\pi^0) / {\cal
B}(D^{*\ }_s\to D_s\gamma)}= 0.036$.  As we will see below, a large $g$
solution with $\chi^2 \lesssim 1$ is only possible if $g$ increases to $\sim
0.9$ and $\beta$ increases to $\sim 6.0\,{\rm GeV^{-1}}$ (c.f. 
Fig.~\ref{fig_rand}).

The experimental limit $\Gamma(D^{*+}) < 0.13\,{\rm MeV}$ \cite{ACCMOR}
translates into an upper bound on the value of $g$.  Since ${\cal B}(D^{*+}\to
D^+\gamma)$ is small, this bound is almost $\beta$ independent and to a good
approximation is
\begin{eqnarray}
  g < 0.52\,\sqrt{\,\sqrt{1+3.01\,x}-1} \qquad\qquad 
     x=\Gamma(D^{*+})^{\:\rm limit}/(0.13\,{\rm MeV}) \,.  \label{ulmt}
\end{eqnarray}
For the situation in row five of Table~\ref{table_solns} this excludes the
hatched region in Fig.~\ref{fig_solns}.  The limit on $\Gamma(D^{*+})$
therefore eliminates the $g\simeq 0.76$ solution at the two sigma level.  Since
this limit has not been confirmed by other groups it would be useful to have
further experimental evidence that could exclude this solution.  

The central values in Eq.~(\ref{soln1}) have uncertainty associated with
the parameter $m_c$.  Taking $m_c=1.4\pm0.1\,{\rm GeV}$ gives $0.25 < g < 0.28$
and $0.79\,{\rm GeV^{-1}} < \beta < 0.93\,{\rm GeV^{-1}}$ for the first
solution, and $0.72 < g < 0.80$ and $4.6\,{\rm GeV^{-1}} < \beta < 5.3\,{\rm
GeV^{-1}}$ for the second solution (in both cases the $\chi^2$ changes very
little).  There is also ambiguity in the solution in Eq.~(\ref{soln1}) due to
the choice of scale $\mu$ (ie., the value of the counterterms $\alpha_1$,
$\alpha_5$, $\tilde\kappa_1$ and $\kappa_5$).  Increasing $\mu$ to $1.3\,{\rm
GeV}$  gives solutions $(g=0.28, \beta=0.91\,{\rm GeV^{-1}}, \chi^2=1.4)$ and
$(g=0.78,\beta=5.0\,{\rm GeV^{-1}},\chi^2=4.1)$, while decreasing $\mu$ to
$0.7\,{\rm GeV}$ gives solutions $(g=0.25, \beta=0.83\,{\rm GeV^{-1}},
\chi^2=3.7)$ and $(g=0.72, \beta=4.7\,{\rm GeV^{-1}}, \chi^2=3.1)$.  Note that
the $\chi^2$ of the second solution remains large, while the $\chi^2$ of the
first solution is reduced significantly by an increased scale.

Another method of testing the effect of the unknown counterterms
$\tilde\alpha_1$, $\alpha_5$, $\tilde\kappa_1$ and $\kappa_5$ is to take their
values at $\mu=1\,{\rm GeV}$ to be randomly distributed within some reasonable
range of values.  We take $-1 < \tilde\kappa_1, \kappa_5 < 1$ and $-2 <
\tilde\alpha_1, \alpha_5 < 2$, with the motivation that the counterterms change
the tree level value of $Z_{\pi}^a$ and $Z_\gamma^a$ by less than $30\%$, and
give corrections that are not much bigger than those from the one-loop graphs. 
Near each of the two solutions $5000$ values of $g$ and $\beta$ were then
generated by minimizing the $\chi^2$.  This gives the distributions in
Fig.~\ref{fig_rand}.  The solution with $g=0.265$ and $\beta=0.85\,{\rm
GeV^{-1}}$ has fairly small uncertainty from the counterterms.  The $g=0.76$,
$\beta=4.9\,{\rm GeV^{-1}}$ solution has much larger uncertainty because the
corresponding contour lines in Fig.~\ref{fig_solns} are almost parallel.  For
this solution the upper bounds are determined by the limits of a few 
${\rm MeV}$ \cite{PDG} on the $D^*$ widths.  
From this analysis we estimate the theoretical uncertainty of the solutions in 
Eq.~(\ref{soln1}) to be roughly
\begin{eqnarray}
 g=0.265^{+0.05}_{-0.02} \ \ \ \beta=0.85^{+0.3}_{-0.1}\:{\rm GeV^{-1}} \:,
    \qquad\quad
 g=0.76^{+0.2}_{-0.1} \ \ \ \beta=4.9^{+5.0}_{-0.7}\:{\rm GeV^{-1}} 
  \label{soln2}
\end{eqnarray}
at this order in chiral perturbation theory.  The errors on $g$ and $\beta$ are
positively correlated since the values of $g$ and $\beta$ are constrained in
one direction by the small error on the $D^{*0}$ rate ratio in 
Eq.~(\ref{exptratios}).  

From Eq.~(\ref{ulmt}) and Fig.~\ref{fig_rand}, we see that if the error in
${{\cal B}(D^{*\ }_s\to D_s\pi^0) / {\cal B}(D^{*\ }_s\to D_s\gamma)}$ can be
decreased by a factor of two, in conjunction with a limit of $\Gamma(D^{*+})
\lesssim 0.6\,{\rm MeV}$ then this could provide strong evidence that the
$g=0.76$ solution is excluded.  On the other hand if the central values of the
second and third ratios in Eq.~(\ref{exptratios}) decrease, then a width
measurement or stronger limit on $\Gamma(D^{*+})$ will be needed to distinguish
the two solutions.

Using the extracted values of $g$ and $\beta$ gives the widths shown in
Table~\ref{table_widths}.  The couplings were extracted at one-loop and order
$m_q \sim 1/m_c$, so the predictions for the $D^*$ widths are made at this
order.  The experimental uncertainty in the $D^*$ widths is estimated by
setting $g$ and $\beta$ to the extremal values in Eq.~(\ref{soln1}), which
gives the range shown in the second and fourth rows of the Table.  The
uncertainty from the unknown counterterms in the third and fifth rows is
estimated in the same way using the uncertainties from Eq.~(\ref{soln2}).  Note
that for the $g=0.265$ solution the $D_s^*$ width is small due to a delicate
cancellation in $\mu_3$ resulting from setting $Z_{\rm wf}^a \times 1/m_c =
1/m_c$.  Keeping $Z_{\rm wf}^a/m_c$ to order $m_q$ gives a $D^*_s$ width of
$0.28\,{\rm keV}$ with a range of $0.1-0.4\,{\rm keV}$ for both the 
experimental and the counterterm uncertainties.

Making use of HQS allows us to predict the width of the $B^*$ mesons from their
dominant mode $B^* \to B\gamma$.  Eq.~(\ref{rateg}) gives the rate for $B^*\to
B\gamma$ with $Q'=-1/3$ and $m_c\to m_b$.  Since the couplings $\beta_1$ and
$\beta_2$ are unknown these rates can not be determined at order $1/m_{c,b}$, but
we can include the order $m_q$ corrections. The $B$ meson masses are taken
from \cite{PDG} and we use $m_b=4.8\,{\rm GeV}$ \cite{glkw}.  We set
$\delta=0.047\,{\rm GeV}$\, and $\Delta=k\cdot v=0$, but since the contribution
$Q'/m_b$ in Eq.~(\ref{rateg}) is numerically important it is kept in our estimate.  For
comparison the widths obtained with the $g=0.76$ and $\beta=4.9\,{\rm GeV^{-1}}$
solution are also shown.  
\begin{table}[!t]
\begin{center}
{\tighten
\begin{tabular}{cccc|ccc}  
 Predicted widths in ${\rm keV}$
 & $D^{*0}$ & $D^{*+}$ & $D_s^{*}$ & $B^{*+}$ & $B^{*0}$ & $B_s^{*}$ \\ \hline 
  $g=0.265$, $\beta=0.85\,{\rm GeV^{-1}}$ & 18 & 26 & 0.06 &
    $\sim$ 0.06 & $\sim$ 0.03 & $\sim$ 0.04 \\
  uncertainty from experiment  & 16 - 24 & 23 - 35 & 0.01 - 0.13\ \  & 
    $-$ & $-$ & $-$ \\
  uncertainty from counterterms  & 16 - 27 & 22 - 39 & 0.04 - 0.13\ \ &
    $-$ & $-$ & $-$ \\ \hline 
  $g=0.76$, $\beta=4.9\,{\rm GeV^{-1}}$ &  323 &  448 & 103 & 
    $\sim$ 2.1 & $\sim$ 2.0 & $\sim$ 1.6 \\
  uncertainty from experiment & 285 - 367 & 396 - 508 & 83 - 128\phantom{7} & 
    $-$ & $-$ & $-$ \\
  uncertainty from counterterms  & 215 - 1318 & 281 - 1157 & 53 - 1078 \ \ &
    $-$ & $-$ & $-$ \\ 
\end{tabular} }
\end{center}
{\tighten \caption{Widths in ${\rm keV}$ for the $D^*$ and $B^*$ mesons.  The
experimental and counterterm ranges are determined by the extremal values of
$g$ and $\beta$ in Eqs.~(\ref{soln1}) and (\ref{soln2}).  For $g=0.265$ the
$D_s^*$ width is small due to a delicate cancellation in $\mu_3$ as
explained in the text. The uncertainty in the $B^*$ widths is large due to
unknown $1/m_{c,b}$ corrections.}  \label{table_widths} }
\end{table} 

As a final comment, we note that heavy meson chiral perturbation theory can
also be used to examine excited $D^{(*)}$ mesons, such as the p-wave states,
$D_0^*$, $D_1^*$, $D_1$, and $D_2^*$ \cite{excited,review}.  To do so, explicit
fields for these particles may be added to the Lagrangian giving a new
effective theory. For interactions without external excited mesons (such as the
ones considered here) these new particles can then contribute as virtual
particles.  However, since we have not included these heavier particles they
are assumed to be `integrated out', whereby such contributions are absorbed
into the definitions of our couplings.

\section{Conclusion}

For the $D^{*0}$, $D^{*+}$, and $D^*_s$, the decays $D^* \to D\pi$ and $D^* \to
D\gamma$ are well described by heavy meson chiral perturbation theory.  Using
the recent measurement of ${\cal B}(D^{*+} \to D^+ \gamma)$ \cite{CLEO}, the
ratios of the $D \gamma$ and $D\pi^0$ branching fractions were used to extract
the $D^*D\pi$ and $D^*D\gamma$ couplings $g$ and $\beta$.  Two solutions were
found
\begin{eqnarray}
 g=0.265\,^{+.04}_{-.02} \,^{+.05}_{-.02}\ \ 
	\beta=0.85^{+.2}_{-.1}\,^{+.3}_{-.1}\,{\rm GeV^{-1}}\, \quad\quad
 g=0.76\,^{+.03}_{-.03} \,^{+.2}_{-.1}\ \ 
 	\beta=4.9^{+.3}_{-.3}\,^{+5.0}_{-.7}\,{\rm GeV^{-1}}\,.
\label{soln_concl}
\end{eqnarray}
The first error here is the one sigma error associated with a minimized
$\chi^2$ fit to the three experimental branching fraction ratios (see
Fig.~\ref{fig_solns}).  The second error is our estimate of the uncertainty in
the extraction due to four unknown counterterms $\tilde\alpha_1$, $\alpha_5$,
$\tilde\kappa_1$ and $\kappa_5$ that arise at order $m_q$ (see
Fig.~\ref{fig_rand}).  

It is possible that the uncertainty from these counterterms can be reduced by
determining them from other processes. For these corrections to contribute at
low enough order in the chiral expansion we need processes with outgoing
photons or pseudo-goldstone bosons, such as semileptonic $D$ decays to $K$,
$\eta$, or $\pi$.  Here there are also SU(3) corrections to the left handed
current which involve an unknown parameter $\eta_0$ \cite{BG}.  Information on
$\kappa_1$ and $\kappa_1'$ can be determined from the pole part of the $D_s\to
K\ell\nu_\ell$ form factor \cite{BG}.  In a similar manner $D_s \to
\eta\ell\nu_\ell$ can constrain $\tilde\kappa_1$ and $\kappa_5$, and a
comparison of the form factors for $D^+\to \bar K^0\ell\nu_\ell$ and $D_s\to
\eta\ell\nu_\ell$ gives information on $\kappa_1'$ and $\kappa_5$.  These
investigations were beyond the scope of this paper. In principle, information
about the constants $\tilde\alpha_1$, and $\alpha_5$ could be obtained from a
measurement of $B\to \gamma\ell\nu_\ell$. The CLEO experimental bound on $B\to
\ell\nu_\ell$ ($\ell=e,\mu$)\cite{CLEO2} is roughly two orders of magnitude
above the theoretical prediction, but due to the helicity suppression for $B\to
\ell\nu_\ell$ the branching ratio for $B\to \gamma\ell\nu_\ell$ may be up to an
order of magnitude bigger\cite{bgenu}.  

Another possible approach would be to use large $N_c$ scaling for the
counterterms in $\delta{\cal L}_g$ and $\delta{\cal L}_\beta$.  Terms that have two
chiral traces are suppressed by a power of $N_c$ compared to those with only one
trace.  In the large $N_c$ limit the counterterms $\tilde\kappa_1$ and
$\tilde\alpha_1$ would dominate, and $\kappa_5$ and $\alpha_5$ could be
neglected, thus reducing the theoretical uncertainty.

The smaller solution for $g$ in Eq.~(\ref{soln_concl}) is fairly insensitive to
the addition of the one-loop corrections (see Table~\ref{table_solns}). 
However, corrections at order $m_q\sim 1/m_c$, including the heavy meson mass
splittings, were important in determining the solution with larger $g$.  The
limit $\Gamma(D^{*+}) < 0.13\,{\rm MeV}$ \cite{ACCMOR} gives an upper bound on
the coupling $g$ (see Eq.~(\ref{ulmt}) and Fig.~\ref{fig_solns}), and
eliminates the $g=0.76$, $\beta=4.9\,{\rm GeV^{-1}}$ solution.  Experimental
confirmation of this limit is therefore desirable.  Note that the largest
experimental uncertainty in our extraction comes from the measurement of ${\cal
B}(D_s^*\to D_s \pi^0)$, and dominates the theoretical uncertainty due to decay
via single photon exchange.  A better measurement of ${{\cal B}(D^{*\ }_s\to
D_s\pi^0) / {\cal B}(D^{*\ }_s\to D_s\gamma)}$ along with a limit
$\Gamma(D^{*+})\lesssim 0.6\,{\rm MeV}$ could provide further evidence that the
$g=0.76$ solution is excluded.  However, if the central values of the second
and third ratios in Eq.~(\ref{exptratios}) decrease then a width measurement or
stronger limit on $\Gamma(D^{*+})$ will be needed to distinguish the two
solutions.  An improved measurement of ${\cal B}(D^{*\ }_s\to
D_s\pi^0)$ may also give valuable information on the unknown couplings
$\tilde\kappa_1$, $\kappa_5$, $\tilde\alpha_1$, and $\alpha_5$.  

The extraction of $g$ has important consequences for other physical quantities
[2-11].  For example\cite{pc}, for the $B\to \pi\ell\bar\nu_\ell$ form factors
with $E_\pi < 2\,m_\pi$, analyticity bounds combined with chiral perturbation
theory give $g\,f_B \lesssim 50\,{\rm MeV}$ \cite{disp}.  The solution
$g=0.265$ gives $f_B \lesssim 190\,{\rm MeV}$ for the $B$ decay constant. 
However, for $g=0.76$ we have $f_B \lesssim 66\,{\rm MeV}$, which is roughly a
factor of three smaller than lattice QCD values, $f_B \simeq 160-205$
\cite{lattice}.

\medskip\noindent {\bf Acknowledgments} \medskip

I would like to thank Zoltan Ligeti and Mark B. Wise for early discussions on
this subject and useful suggestions.  I would also like to thank Aneesh
Manohar, Tom Mehen, Vivek Sharma, Hooman Davoudiasl, and Martin Gremm for
helpful comments.  This work was supported in part by the U.S.\ Dept.\ of
Energy under Grant no.\ DE-FG03-92-ER~40701.  

\appendix
\section*{One loop correction formulae}

In this appendix we give explicit formulas for the functions $G_1$, $F_1$, 
$F_2$, $\varrho_\pi^a$, and $\varrho_\gamma^a$ that occur in our one loop 
correction formulae in Eqs.~(\ref{wfnren}), (\ref{pi12vertex}), 
(\ref{pi3vertex}), and (\ref{vertexg}).  In doing this type of one-loop 
calculation an important integral is
\begin{eqnarray}
   \int {d^{4-2\epsilon}q \over (2\pi)^{4-2\epsilon} } { \mu^{2\epsilon}
 \over (q^2-m^2+i\varepsilon)\,2( q\cdot v - b + i\varepsilon) } = 
 -{i\, b \over (4\pi)^2} \left[ \frac1{\hat\epsilon} +\ln{(\frac{\mu^2}{m^2})} 
 + 2 - 2 F(\frac{m}{b})  \right] \,, \label{nastint}
\end{eqnarray}
where $1/\hat\epsilon=1/\epsilon-\gamma+\log{(4\pi)}$.
$F$ is needed for both positive and negative $b$, so 
\begin{eqnarray}
 F\left(\frac1{x}\right) &=& \left\{ \begin{array}{l}
  -\mbox{\large ${\sqrt{1-x^2}\over x}$} \left[\mbox{\large $\frac{\pi}2$} 
  -\tan^{-1}\left( \mbox{\large $\frac{x}{\sqrt{1-x^2}}$} \right) \right] 
  \quad\qquad |x|  \le 1 \\
  \phantom{-} \mbox{\large ${\sqrt{x^2-1}\over x}$}\: \ln{\left(x+\sqrt{x^2-1} 
  \,\right)} \quad\qquad\quad\ \ \ |x| \ge 1 \end{array} \right.\,. \label{F}
\end{eqnarray}
For $b>0$ the function $F$ was derived in \cite{manjen,FG} and agrees with the
above formula{\tighten\footnote{Eq.~(\ref{F}) for $F$ disagrees with
\cite{BG} for $x<0$.  Their $F(1/x)$ is even under $x\to -x$ making
Eq.~(\ref{nastint}) discontinuous at $\Delta=0$.  Furthermore, their $F$ has no
imaginary part corresponding to the physical intermediate state.}}.  For
$x=b/m < -1$ the logarithm in Eq.~(\ref{F}) has an imaginary part. This
corresponds to the physical intermediate state where a heavy meson of mass
$m_H$ produces particles of mass $m_H+b$ and $m$.  For the calculation
here the imaginary part only contributes from $F(m_\pi/(d_0-\Delta))$, and was
found to always be numerically insignificant.  Note that the real part of
$x\,F(1/x)$ is continuous everywhere, and differentiable everywhere except
$x=-1$.  Also $F(1)=F(-1)=0$.

Eq.~(\ref{wfnren}) contains the function
\begin{eqnarray}
 G_1(a,b) &=& \frac53\,a^2 + (4b^2-\frac43 a^2) F(a/b)
   + \frac43 (a^2-b^2) \frac{a}{b} F'(a/b) \,, \label{G1G2} 
\end{eqnarray}
where $a^2$ is the analytic contribution.  In the limit $\Delta\to 0$
Eq.~(\ref{wfnren}) gives $Z_{D}=Z_{D^*}$ in agreement with HQS.  To obtain
HQS in the finite part of the dimensionally regularized calculation of the
graphs in Fig.~\ref{fig_wf} it was necessary to continue the $D^*$ fields to
$d=4-2\epsilon$ dimensions (so the $D^*$ polarization vector
$\epsilon_\alpha=(1-\frac{\epsilon}3) \tilde\epsilon_\alpha$ where $\sum
\tilde\epsilon_\alpha^{\,*} \tilde\epsilon^\alpha = -3$).  

In Eqs.~(\ref{pi12vertex}) and (\ref{pi3vertex}) we have the functions
\begin{eqnarray}\label{varrhopi}
  F_1(a,b,c) &=& -\frac76 \,a^2 + \frac{2}{3(b-c)}\left[
   b(a^2-b^2)F(a/b) - c(a^2-c^2)F(a/c) \right] \,,  \nn \\
 \varrho_{\pi}^{a=1,2} &=&  {m_\pi^2\over (4\pi f)^2} {\tilde\kappa_1(\mu) 
   \over 2}  \,, \\
 \varrho_{\pi}^3 &=& {1\over (4\pi f)^2}\: {(m_u-m_d)\over 2(m_s-\hat m)}\: 
    \bigg[ (m_K^2-\frac{m_\pi^2}2)\,\tilde\kappa_1+(m_K^2-m_\pi^2)\,\kappa_5
    \bigg] + {(m_{K^\pm}^2-m_{K^0}^2) \over (4\pi f)^2}\:  \kappa_5 \,. \nn
\end{eqnarray}
We have ignored isospin violating counterterm corrections in 
$\varrho^{1,2}_\pi$ and work to leading order in the isospin violation for
$\varrho^3_\pi$.  In deriving Eq.~(\ref{varrhopi}) use has been made of 
$m_\pi^2 = 2B_0 m_u = 2B_0 m_d =2B_0 \hat m$, $m_K^2-m_\pi^2/2=B_0 m_s$, 
and $m_{K^\pm}^2-m_{K^0}^2=(m_u-m_d) B_0$.

In Eq.~(\ref{vertexg}) we have the function $F_1$ and the function
\begin{eqnarray}
 F_2(a,b,c) &=& -2\,b-c-\frac{2a^2}{c}\int^{a/(b+c)}_{a/b} dt {F(t)\over t^3} 
  \nn\\
  &=&-2\,b-c- \frac{2a^2}{c} \left[ { 1\over 4x^2 }- {F(x)\over 2\,x^2} 
  - {F(x)^2 \over 4\,(x^2-1) } \right] \Bigg|_{x=a/b}^{x=a/(b+c)} \,.
\end{eqnarray}
Assuming isospin to be conserved the counterterm contributions in 
Eq.~(\ref{vertexg}) are
\begin{eqnarray}
 \varrho_{\gamma}^{\,a=1,2} &=& {m_\pi^2\: \tilde\alpha_1 \over 
  2\,(4\pi f)^2}\: - {(m_K^2-m_\pi^2)\:\alpha_5 \over 3\,Q_{aa}\,(4\pi f)^2} 
  \,,\nn\\
 \varrho_{\gamma}^3 &=& {(2\,m_K^2-m_\pi^2)\:\tilde \alpha_1 \over 
  2\,(4\pi f)^2} +{(m_K^2-m_\pi^2)\:\alpha_5\over (4\pi f)^2} \,. 
\end{eqnarray}

{\tighten

} 

\end{document}